\documentclass{aa}  
\usepackage{graphicx}

\usepackage{natbib}
\bibpunct{(}{)}{;}{a}{}{,} 

\usepackage[hidelinks]{hyperref}
\usepackage{lscape}
\usepackage{dirtytalk}
\hypersetup{
    colorlinks=true,
    linkcolor=blue,
    filecolor=blue, 
    citecolor=blue,
    urlcolor= blue,
    }
\usepackage{txfonts}

\newcommand{\msun}{$M_{\rm \odot}$}
\newcommand{\lsun}{$L_{\rm \odot}$}

\newcommand{\zsun}{$Z_{\rm \odot}$}

\usepackage[dvipsnames]{xcolor}

\makeatletter
\renewcommand*\aa@pageof{, page \thepage{} of \pageref*{LastPage}}
\makeatother

\begin{document}

   \title{On the evolutionary nature of puffed-up stripped star binaries and their occurrence in stellar populations }
 
\author{Debasish Dutta
          \inst{1}
          \and
         Jakub Klencki\inst{2}
          }

   \institute{Department of Physics, Indian Institute of Technology Palakkad, Kerala, India\\
              \email{debasishdutta1707@gmail.com}
         \and
              European Southern Observatory, Karl-Schwarzschild-Strasse 2, 85748 Garching bei München, Germany\\
                \email{Jakub.Klencki@eso.org}
               }

   \date{Received XXXX; accepted XXXX}

 
  \abstract
  {The majority of massive stars are formed in multiple systems and at some point during their life they interact with their companions via mass transfer. This interaction typically leads to the primary shedding its outer envelope, resulting in the formation of a "stripped star". Classically, stripped stars are expected to quickly contract to become hot and UV-bright helium stars. Surprisingly, recent optical spectroscopic surveys have unveiled a large number of stripped stars that are much larger and cooler, appearing "puffed up", and overlapping with the Main Sequence (MS).nature of the phase in which stripped stars contract.  Here, we study the evolutionary nature of puffed-up stripped (PS) stars and the duration of this enigmatic phase using stellar-evolution code MESA. We computed grids of binary models at four metallicities: Solar (Z$_{\odot}$ = 0.017), Large Magellanic Cloud (LMC, Z=0.0068), Small Magellanic Cloud (SMC, Z = 0.0034), and Z = 0.1Z$_\odot$. Contrary to previous assumptions, we find that stripped stars regain thermal equilibrium shortly after the end of mass transfer and maintain it during most of the PS phase.Their further contraction towards hot and compact He stars is determined by the rate at which the residual H-rich envelope is depleted, with the main agents being H-shell burning (dominant for M $\lesssim$ 50 \msun)
  and mass-loss in winds. 
  The duration of the PS star phase is approximately 10\% of the core-He burning lifetime (1\% total lifetime) and up to 100 times more than thermal timescale. It decreases with increasing mass and luminosity and increases with metallicity. We explored several factors that influence PS star lifetimes:  orbital period, mass ratio, winds, and semiconvection. We further carried out a simple binary population synthesis estimation, finding that $\sim$0.5 – 0.7\% of all the stars with log (L/L$_\odot$) > 3.7 may in fact be PS stars. Our results indicate that tens to hundred of PS stars in post-interaction binaries may be hiding in the MS population, disguised as 'normal' stars: $\sim$100 ($\sim$280) in the SMC (LMC) alone. Their true nature may be revealed by spectroscopically measured low surface gravities, high N enrichment, and likely slow rotation rates.}


    \titlerunning{Evolutionary nature of puffed-up stripped star binaries }   
  \maketitle
  %

\section{Introduction}

Binary stars form a significant portion of the stellar population. Approximately 40\% of low-mass stars (0.8-1.2\msun) exist within binary or multiple systems \citep{2017ApJS..230...15M}. This prevalence is even more pronounced among massive stars (M $\gtrsim$ 8 \msun), where binary systems overwhelmingly dominate the overall population \citep{2012Sci...337..444S, 2014ApJS..215...15S}. Around 94\% of massive stars with masses $\gtrsim$ 16 $M_{\rm \odot}$ exist within binary or multiple systems \citep{2017ApJS..230...15M}, where most of them engage in interactions with companion stars, involving the transfer of mass and angular momentum. During mass transfer, the more massive star, often referred to as the primary star, transfers material to its less massive companion through a process known as Roche Lobe Overflow (RLOF). This can lead to the primary star shedding its outer envelope, resulting in the formation of what is known as a "stripped star". A large fraction of massive binaries ($\sim30\%$) go through envelope stripping \citep{2012Sci...337..444S} and therefore are more common \citep{1992ApJ...391..246P, 2013MNRAS.436..774E,2019A&A...629A.134G,2021ApJ...908...67S}. Due to their very high temperatures, stripped stars serves as a important source for ionizing photons and are considered to have had an important contribution to cosmic reionization \citep{2018A&A...615A..78G, 2020A&A...634A.134G}. They are recognized as progenitors of stripped envelope supernovae \citep{2011MNRAS.412.1522S,2017ApJ...840...10Y} and are also deemed to be one of the key contributors to the population of gravitational wave progenitors, leading to the formation of merging pairs of neutron stars (NSs) and black holes (BHs) \citep{2007PhR...442...75K, 2020ApJ...904...56G}.

Stripped stars are most commonly formed when the primary star donates mass to its companion during stable mass transfer initiated as a result of rapid expansion after the end of the main sequence (crossing the Hertzsprung gap). During this process, the donor star rapidly loses its entire hydrogen envelope during a short-lived phase of thermal timescale mass exchange ($\lesssim$ $10^4$ years) \citep{1967ZA.....65..251K,1975ApJ...198L.109V,1992ApJ...391..246P,1998A&ARv...9...63V} (see however, \citealp{2022A&A...662A..56K}, where it is shown that the mass transfer can also occur at nuclear timescale in case of low metallicity massive donors). This kind of mass transfer is referred to as case B mass transfer \citep{1967ZA.....65..251K, 1967AcA....17..355P, 2001A&A...369..939W} and is the most common channel for mass transfer of stars in binaries \citep{2008AIPC..990..230D, 2017A&A...608A..11G, 2021ApJ...916L...5V}. Stripped stars can also be formed when stable mass transfer occurs during the main sequence (referred to as case A; \citep{2001A&A...369..939W}), or after the ejection of a common envelope formed as a result of unstable mass transfer \citep{1967ZA.....65..251K, 2011ApJ...730...76I}. Stars undergoing mass loss through these processes often result in the formation of stripped stars of various kinds, contributing to a diverse range from subdwarfs to Wolf Rayet (WR) stars \citep{1967AcA....17..355P,1975ApJ...198L.109V,1991A&A...252..159V,1992ApJ...391..246P,1992A&AS...94..453D,2005A&A...435.1013P,2008MNRAS.384.1109E,2017A&A...608A..11G,2020A&A...637A...6L}. Subdwarfs are stars with masses up to about 1.5 \msun, which correspond to initial masses of up to about 7 $M_{\rm \odot}$ and account for low mass stripped stars. WR stars have masses $\gtrsim$ 8 (although this depends strongly on metallicity, see for example \citealp{2020A&A...634A..79S}) and accounts for the massive stripped stars.  Intermediate mass stripped stars mostly refers to stars with mass between $\sim$1 to 8 $M_{\rm \odot}$ (progenitor masses between 8 and 25 \msun).

After the mass transfer, these stars subsequently transition into hot, compact helium stars \citep[eg.][]{1969A&A.....3...83K,2017A&A...608A..11G,2018A&A...615A..78G,2020A&A...637A...6L, 2021A&A...656A..58L,2022A&A...662A..56K}. This phase spans about 10\% of the star's entire lifespan \citep{2017A&A...608A..11G,2018A&A...615A..78G}, with its main source of energy arising from helium fusion through the triple alpha reaction occurring in the core. However, the majority of emitted light during this stage primarily occurs in the ultraviolet (UV) range, rendering it inaccessible to observation by  ground-based telescopes \citep{2017A&A...608A..11G}. This region is characterized by very high temperature and populated by subdwarf O (sdO) stars and Wolf-Rayet (WR) stars \citep[see][]{2017A&A...608A..11G,2018A&A...615A..78G}. A handful of such systems are observed till date, including sdO stars like $\phi$ Persei \citep{1998ApJ...493..440G}, 59 Cyg \citep{2013ApJ...765....2P}, and 60 Cyg \cite{2017ApJ...843...60W}, along with a quasi-Wolf-Rayet (WR) star identified in HD 45166 \citep{2005A&A...444..895S,2008A&A...485..245G,2023Sci...381..761S}. Additionally, sample of Galactic sdOs  have been reported by \cite{2018ApJ...853..156W, 2021AJ....161..248W}. Notably, \cite{2023arXiv230700061D} have recently reported a sample of 25 stars situated in the Magellanic Clouds. These stars exhibit characteristics in line with the expected properties of intermediate mass stripped stars, encompassing mass ranges from 1 to 9 \msun, as discussed in \cite{2023arXiv230700074G}.

Interestingly, recent observations have detected systems which have recently detached from mass transfer and have not yet settled as a hot and compact helium star. These stars are characterized by larger radii and cooler effective temperatures than classical \lq hot' stripped stars and are prevalent as "bloated stripped stars" in recent literature \citep{2020A&A...641A..43B, 2022MNRAS.511L..24E,2023MNRAS.525.5121V}, etc.). They are visible in the optical spectrum due to their relatively cooler surface temperatures than the hot stripped stars. However, they pose a challenge in terms of accurate identification, as they resemble normal B type stars with their HR diagram position (temperatures and luminosities) often overlapping with the Main Sequence (MS). Their true nature is revealed in spectroscopy when the surface abundances (in particular enhanced N) and relatively small gravities (leading to low spectroscopic masses) are measured. However, this measurement is not an easy task since they are usually accompanied by a rapidly-rotating companion with a similar spectral type, making it difficult to disentangle the cool stripped star from the combined spectra of the binary without sufficient spectral resolution and multi-epoch data.

One famous example is LB-1, which was reported by \cite{2019Natur.575..618L} as a $\sim$70 $M_{\rm \odot}$ black hole (BH). However, \cite{2020A&A...639L...6S} subsequently presented an alternative interpretation, suggesting that the system could be better explained as comprising a \lq puffed-up' stripped star with a mass of 1.5 \msun, accompanied by a Be companion weighing 7 \msun.  Another example is HR 6819, which was proposed as a triple system containing a BH in the inner binary by \cite{2020A&A...637L...3R}. However, \cite{2020A&A...641A..43B} (also \citealp{2021MNRAS.502.3436E}) found it to contain a \lq puffed-up' stripped star of mass 0.56 \msun. Similar ambiguity surrounds NGC 1850 BH1, initially proposed as an 11 \msun BH by \cite{2022MNRAS.511.2914S} but later suggested to be a \lq puffed-up' stripped star with a mass between 0.65 and 1.5 $M_{\rm \odot}$ \cite{2022MNRAS.511L..24E}. 

The examples stated above involve low-mass stripped stars (M\textsubscript{stripped} $\lesssim$ 1.5). Among, intermediate mass stripped stars (M\textsubscript{stripped} $\sim$ 1-8 \msun), is the $\gamma$ Columbae, which was reported by \cite{2022NatAs...6.1414I} as a partially stripped pulsating core ($\sim$ $4-5$ \msun) of a massive star of roughly 12 $M_{\rm \odot}$ at solar metallicity. However, no companion was detected. Recently, \cite{2023A&A...674L..12R} discovered the first stripped star of intermediate-mass in low metallicity, SMCSGS-FS 69. This star was found to be a puffed-up stripped star ($\sim$ 3 \msun) originating from a massive star of $\sim$12 $M_{\rm \odot}$ and is currently transitioning towards the hotter part in the HR diagram. Among the other recently discovered ones are  VFTS 291 ($\sim$ 1.5–2.5 \msun) \citep{2023MNRAS.525.5121V} and the so called star 16 ($\sim$ 0.8 \msun) from the recently reported  sample of 25 stars by \cite{2023arXiv230700061D} and \cite{2023arXiv230700074G}.

The phase of evolution where these puffed-up stripped stars are found ranges from post-RLOF to the moment where the companion outshines the primary in the optical spectra/band. Classically, this immediate post-RLOF phase of contraction is perceived to be a short-lived temporary stage, with the duration likely characterized by the thermal timescale of the stripped donor ($\lesssim$10$^4$ yr). This, however, would correspond to only < 0.1\% of the total stellar lifetime, most likely at odds with the number of puffed-up stripped stars detected so far. This conundrum was noted by \cite{2021MNRAS.502.3436E}, who briefly pointed out that the duration of the post-RLOF contraction and the puffed-up stripped star phase in their models are much longer than the thermal timescale. A comprehensive theoretical investigation focused on this specific phase of binary star evolution is, however, still missing.

Consequently, this paper endeavors to fill this gap by offering an in-depth examination of the immediate post-RLOF phase. Our focus centers on unraveling the intricate evolution of these puffed-up stripped stars. For this purpose, we generate grids of binary models between 5 and 40 $M_{\rm \odot}$ at four distinct metallicities: solar metallicity (Z = 0.017), Large Magellanic Cloud (LMC) metallicity (Z = 0.0068), Small Magellanic Cloud (SMC) metallicity (Z = 0.0034) and Z = $0.1Z_{\odot}$ = 0.0017. We also investigate several factors that could potentially impact the duration of the contraction phase,  such as orbital period, mass ratio, wind mass loss rate and semiconvection. Furthermore, we also estimate the number of puffed-up stripped stars one can expect to find in the small and large magellanic clouds.

The structure of the paper is as follows: In Sec. \ref{sec2}, we outline our computational method and the physical assumptions underlying our study, in Sec. \ref{sec3}, we perform a detailed investigation of one particular binary model, and further discuss the full grids of evolutionary models. In Sec. \ref{sec4}, we discuss the  we discuss the uncertainties and estimate the number of PSS that we predict to exist in young stellar populations and compare our default definiton of the puffed-up stripped star phase with another criteria.  We summarise and conclude our findings in Sec. \ref{sec5}.

\section{Method: Binary stellar evolution models} \label{sec2}
\subsection{Physical ingredients}
We construct binary evolution models using the 1D stellar evolution code MESA \citep{2011ApJS..192....3P, 2013ApJS..208....4P, 2015ApJS..220...15P, 2018ApJS..234...34P, 2019ApJS..243...10P} (version 11540). Models were computed using four distinct initial chemical compositions. For solar metallicity, we assumed $Z_\odot$ = 0.017 and abundance ratios as outlined by \citep{1998SSRv...85..161G} \citep[also][]{1996ASPC...99..117G}. In case of sub solar metallicities, we computed our models using abundance ratios of the Magellanic Clouds following \cite{2011A&A...530A.115B}, where we used Z= 0.0068 (0.4 $Z_\odot$) for Large Magellanic Clouds (LMC), Z = 0.0034 (0.2 $Z_\odot$) for the Small Magellanic Clouds (SMC). Additionally, we also computed our models using Z =0.0017 (0.1 $Z_\odot$). All of our models are non-rotating. We employ the standard \texttt{basic.net} nuclear network, comprising 8 distinct species. Additionally, we make use of the default MESA opacity tables in our analysis \citep{1993ApJ...412..752I, 1996ApJ...464..943I,1976ApJ...210..440B, 2007ApJ...661.1094C}.

Convection is modelled as per the mixing length theory \citep{1958ZA.....46..108B} with a mixing length of 2.0 times the pressure scale height. We adopted the Ledoux criterion for convection and accounted for semiconvective mixing with an efficiency parameter of $\alpha_{SC}$ = 0.01. While $\alpha_{SC}$ has limited effect on main sequence (MS) donors, its effect on post MS is enormous \citep{2020A&A...638A..55K}. We studied the effect of high semiconvection in Appendix \ref{appB}. We accounted for convective core-overshooting by applying exponential overshooting \citep{2000A&A...360..952H}, with a parameter $f$ = 0.0425. It is important to note that convective motion comes to a halt at the boundary of a convective zone. However, to account for the phenomenon of exponential overshooting beyond this boundary we consider an additional parameter, $f_0$ = 0.001. We also took into account the thermohaline mixing \citep{1980A&A....91..175K} with thermohaline coefficient = 2.0. However, earlier studies have shown that thermohaline mixing have little impact on the stellar structures of stripped stars \citep{2017ApJ...840...10Y, 2017A&A...608A..11G}.To help converge models approaching the Eddington limit, we applied reduction of superadiabacity in radiation-dominated regions \citep[Sec. 7.2 of][]{2023ApJS..265...15J}.

To compute the interaction involving a binary system, we employ the implicit mass-transfer approach proposed by \cite{1990A&A...236..385K} for the Roche-lobe overflow. The main goal of this paper is to analyze the behavior of the donor star following the RLOF, aiming to comprehend the extended duration of the subsequent contraction phase. Therefore, for simplicity, we exclusively track the evolutionary path of the primary (donor) star, treating the companion as a point mass equal to 0.6 of the initial primary mass. we assume the accretion efficiency of 100\%. Our focus primarily revolves around stripped-envelope stars generated through mass transfer from stars that fill their Roche lobes due to swift expansion during the hydrogen shell burning phase after departing from the main sequence. A minority of cases also involve consideration of case A mass transfer. Furthermore, we incorporate the influence of tidal forces using the methodology outlined by \cite{1981A&A....99..126H}. It is important to note that tidal effects are not significant in massive case B binaries.

We modelled stellar winds following the default "Dutch" wind scheme of MESA \citep{2011ApJS..192....3P}.  This approach for massive stars combines findings from various studies, mostly done by people from the Netherlands. The specific amalgamation incorporated within MESA derives from the work of \cite{2009A&A...497..255G}. This scheme uses the mass loss rates of \cite{2001A&A...369..574V} for T\textsubscript{eff} $\geq$ $10^4$ \& surface H $>$ 0.4 (hydrogen-rich) and the rates of \cite{2000A&A...360..227N} for T\textsubscript{eff} $\geq$ $10^4$ \& surface H $>$ 0.4 (hydrogen-poor). \cite{2001A&A...369..574V} includes a metallicity dependence of $\dot M$ $\propto$ $(Z/Z_\odot)^{0.85}$ and in the case of \cite{2000A&A...360..227N}, it is $\dot M$ $\propto$ $Z^{0.47}$. The Reimers\_scaling\_factor for cool stars and Dutch\_scaling\_factor for hot stars are both set to 1.

\subsection{Initial parameter space and stopping conditions}
\begin{figure*}[ht!]
    \centering
    \includegraphics[width=19cm]{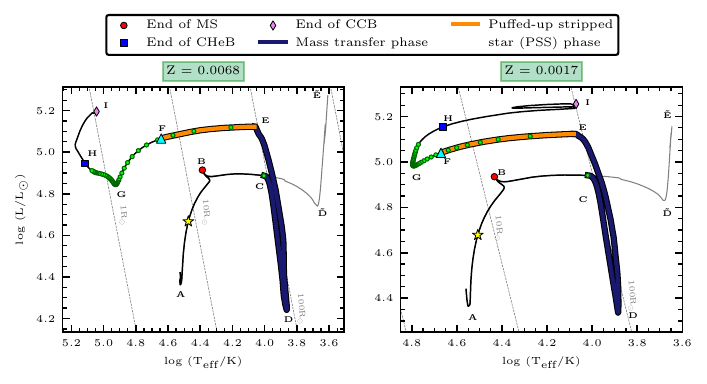}
    \caption{Hertzsprung-Russel diagram of two representative binary models of 16 M$_\odot$ at metallicities 0.0068 (LMC) and 0.0017(0.1\zsun) starting from main sequence (MS) till core carbon depletion. Dotted diagonal lines represent lines of constant radius. The evolution stages are marked in various color and labelled. The red dot represents the end of main sequence (MS), the blue square and pink diamond represents the end of core He burning (CHeB) and core carbon burning (CCB) respectively. The blue region represents the mass transfer phase and orange region shows the puffed-up stripped star (PSS) phase. Leters A to I marks the important instances in the stars evolution (see text). The grey track ($ABC\Tilde{D}\Tilde{E}$) shows the evolutionary pathway of the stars had they not been in a binary configuration (i.e. single stellar evolution). The green dots are placed at an interval of 20,000 years starting from RLOF. The cyan triangle represents the position where the flux of the primary in a HST 450 optical filter drops below the flux of the companion.  For this example we assumed that the companion has the same luminosity and temperature as the primary had at 75\% of its MS lifetime (as indicated by the yellow star, see Sec.\ref{sec4.2} for further discussion)}
    \label{fig:represent}
\end{figure*}

Models are computed with four different chemical compositions: solar (\zsun), LMC (0.4\zsun), SMC (0.2\zsun), 0.1\zsun. Each grid spans 14 different initial primary masses $M_1$: 5, 6.5, 8, 10, 12, 16, 16.9, 20, 22, 26, 30, 32, 36 and 40 M$_\odot$.  The default initial mass ratio $q_{\rm ini}$ = $M_2/M_1$ and the default initial orbital period $P_{\rm ini}$ is 0.6 and 70 days respectively. We have computed additional models encompassing various mass transfer sequences, where q ranges from 0.4 to 1.0, and the initial orbital period $P_{\rm ini}$ spans from 10 to 100 days. We have terminated our models at central Helium depletion (Y $<$ $10^{-6}$). For demonstration purpose, we have terminated some of our models (Fig. \ref{fig:represent} and \ref{fig:nowinds_rxn}) at core Carbon depletion ($X_C$ $<$ $10^{-6}$). The puffed up stripped star phase is considered from the immediate post-RLOF moment until the effective temperature is 0.1 dex higher than the of the Zero Age Main sequence (ZAMS) temperature of the primary. We discuss this choice and compare it to a different criteria further in Sec. \ref{sec4.1}.

\section{Results} \label{sec3}
\subsection{Exploring Evolutionary Models} \label{sec3.1}
In this section, we delve into our findings by closely examining binary models characterized by initial primary mass of 16 $M_{\rm \odot}$  along with initial orbital periods $P_{\rm ini}$ = 70 days and mass ratio $q_{\rm ini}$ = 0.6 . These models are examined under two distinct metallicities: LMC (Z = 0.4 $Z_{\rm \odot}$ = 0.0068) metallicity and Z = 0.0017. Fig. \ref{fig:represent} shows the HR diagram of the primary both cases. The blue lines represent the RLOF phase and the orange lines represent the puffed-up stripped star phase, defined as the stage between the end of the RLOF and the point when the stripped donor becomes hotter than its ZAMS position plus 0.1 dex in $\log T_{\rm eff}$. As we discuss further, this ad hoc temperature threshold approximates the regime when the stripped star is still comparably bright in the optical bands to its likely companion and can most easily be detected without resorting to UV.

To comprehend the mechanisms governing the evolution of the star, we analyze the important factors. In Fig. \ref{fig:panel} (also see Fig. \ref{fig:panel3} and \ref{fig:panel6}), a magnified view of various parameters during the contraction phase is presented. The lowest panel (d) displays the different luminosities contributing to energy output. Notably, the dashed red line ($L_{\text{grav}}$) represents the amount of energy per unit of time related to changes in the gravitational energy of the star. This luminosity is the integral of gravitational energy generation rate per unit mass $\epsilon_{\text{grav}}$, where $\epsilon_{\text{grav}}$ is a local quantity. When $L_{\text{grav}}$ is positive, it signifies that the star is undergoing self-contraction due to its own gravity. Consequently, when $L_{\text{grav}}$ is positive than part of the surface luminosity of the star is due to rapid structural changes (e..g core contraction). Conversely, $L_{\text{grav}}$ being negative indicates that its surface luminosity is lower than that produced in nuclear burning as part of the nuclear energy is used to increase the gravitational potential energy of the star.

In the following sections, we describe the important stages in the evolution of the primary star starting from the main sequence till the end of core Helium burning. For an exemplary choice, we take the 16 $M_{\rm \odot}$ model at LMC metallicity (Z = 0.0068) and also compare it with the lower metallicity model (Z = 0.0017).

\subsubsection{Evolution until RLOF}

\begin{figure}[ht!]
    \centering
    \hspace{-1.1cm}
    \includegraphics[width=10cm]{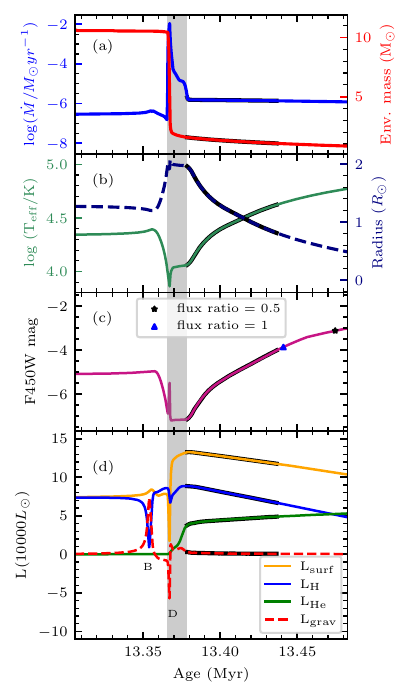}
    \caption{Evolution of various quantities of a 16 M$_\odot$ model (at LMC metallicity) with star age, near the mass transfer phase leading up to the puffed-up stripped star (PSS) phase. Panel (a) shows the variation of envelope mass and mass loss rate of the donor. The mass loss is mainly due to transfer of mass to the companion, but also contains contribution from winds. Panel (b) shows effective temperature (green) and donor radius (blue) and panel (c) shows the magnitude computed in the F450W optical filter of HST WFPC2 camera. The blue triangle and the black star represents the position where the flux ratio ($f_{\rm stripped}/f_{\rm companion}$) drops below 1 and 0.5 respectively. Panel (d) represents various luminosities contributing to the total energy output: the orange line represents the surface luminosity ($L_{\rm surf}$), the blue and green line reprents the luminosity to hydrogen burning ($L_{H}$) and helium burning ($L_{He}$) respectively. The luminosity in red dashed line is the luminosity due to gravitational contraction/expansion ($L_{\rm grav}$). The grey shaded region represents the mass transfer phase and the regions with dark lines represent the puffed-up stripped star (PSS) phase.}
    \label{fig:panel}
\end{figure}

The main-sequence evolution of our primary stars (labeled A-B in Fig \ref{fig:represent}) closely follows the evolution of a single star. The star burns hydrogen into helium in a convective core, with the low-metallicity model being slightly more compact and hotter during the MS due to smaller opacity. After the end of MS (B), the core contracts while the outer layers expand. In Fig. \ref{fig:panel}, we see that $L_{\rm grav}$ increases to nearly equal to the surface luminosity ($|L_{\rm grav}|$ / $L_{\rm surf}$ $\sim$ 0.8), which means that most of the energy is generated from contraction within the star and the shell burning has not kicked in yet. This coincides with a slight increase in $T_{\rm eff}$, a very brief phase just before point B. As a result of this contraction of the star, the shell gets hot and $L_H$ quickly increases, which in turn causes the outer layers (envelope) to expand and the star moves to cooler effective temperatures. The H shell burning starts pretty much straight after the end of the MS and it is this increase in $L_H$ that puffs up the star \citep{2022MNRAS.512.4116F}.

The star continues to expand on a thermal timescale until it fills its Roche lobe, prompting the transfer of matter to its companion. If the star had not been in a binary configuration, it would have followed the trajectory in grey color ($ABC\Tilde{D}\Tilde{E}$) (Fig. \ref{fig:represent}). During mass transfer, as the donor loses a significant amount of mass from its outer layers, a significant fraction of its burning luminosity is trapped in the expanding envelope and used to readjust the stellar structure ( |$L_{\rm grav}$| / $L_{H}$ $\sim$ 0.77 when the ML rate  peaks, see Fig \ref{fig:panel}). During the entire mass transfer phase lasting approximately 0.013 Myr the donor sheds most of its H-rich envelope (nearly 85\% of the envelope mass it had before RLOF), leaving behind only $\sim$1.6 $M_{\rm \odot}$  of the envelope. Notably, the energy generation in shell burning ($L_{H}$) is only weakly affected by the interaction and $L_{H}$ returns to a value even greater than its pre-RLOF  value (see Fig. \ref{fig:panel})
\begin{figure}[ht!]
    \centering
    \includegraphics[width=9cm]{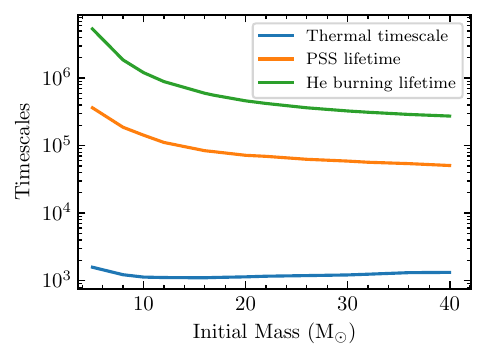}
    \caption{Variation of various timescales with initial mass at LMC metallicity. The blue line represents thermal timescale and the orange line represents the puffed-up stripped star (PSS) lifetime. As can be seen the PSS lifetime is $\sim$100 times longer than the thermal timescale. It is more closer to the Helium burning lifetime which is represented by a green line.}
    \label{fig:timescales}
\end{figure}

\begin{figure}[ht!]
    \centering
    \includegraphics[width=9cm]{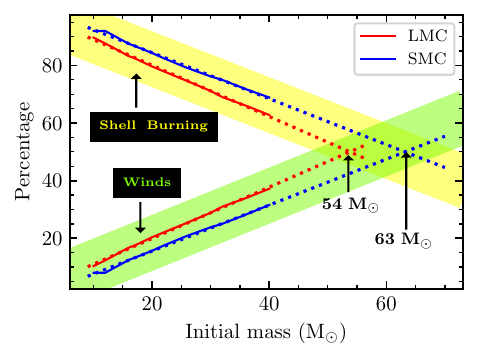}
    \caption{Variation of relative contribution of shell burning and winds to the hydrogen envelope depletion rate during the PSS phase, as a function of the initial donor mass. The yellow parts represent the contribution of shell burning and the green region represents contribution of winds. Red and blue lines represent LMC and SMC composition respectively. It is evident that the shell burning contributes more to the mass depletion of the envelope than the winds. The lines are extrapolated (dotted) to see the extent to which shell burning dominates. In case of LMC, it is $\sim$ 54 $M_{\rm \odot}$ and in case of SMC, it is 63 \msun. Above this limit, the winds start to dominate the mass depletion of the envelope.}
    \label{fig:percent}
\end{figure}

\begin{figure*}
    \centering
    \includegraphics[width=19cm]{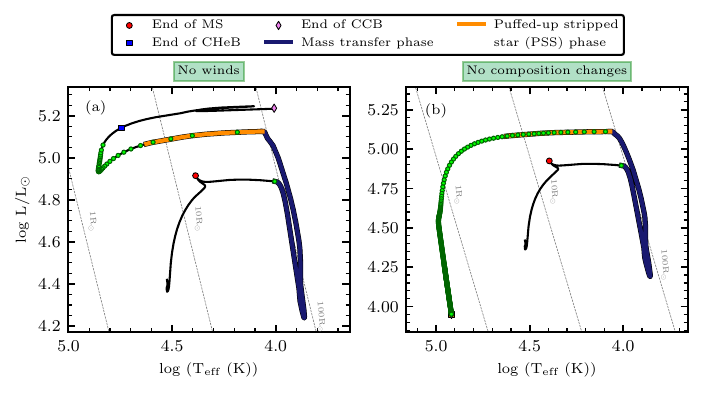}
    \caption{Evolution of 16M$_\odot$ star in LMC composition considering no winds (left) and no composition changes (right). The colors and labels are same as that of Fig. \ref{fig:represent}. For 16\msun, the winds play a subdominant role in stripping the envelope, and completely switching them off only prolongs the PSS phase by $\sim$ 1.3 times. On the other hand, in the hypothetical scenario with no composition changes due to burning (in particular H shell burning), the PSS phase would be much longer, by $\sim$ 4.7 times, illustrating the key role of shell burning in determining the evolution through the PSS phase. The green dots appears so close in the right plot, showing that the contraction happens very slowly as compared to the default scenario.}
    \label{fig:nowinds_rxn}
\end{figure*}

\subsubsection{The "contraction phase"}
Once the mass transfer ceases, we are left with a helium rich star ($M_{\rm core}$ $\approx$ 5.4 \msun, $X_{\rm He; center}$ $\approx$ 0.99) encased in a small hydrogen envelope of $\sim$1.6 \msun. The question of how much envelope mass is left unstripped after the RLOF is closely connected to factors such as metallicity, internal-envelope structure, and the wind mass loss rate. At lower metallicities, stars retain a larger part of the H-rich envelope after the mass transfer \citep{2017A&A...608A..11G, 2017ApJ...840...10Y,2020A&A...637A...6L, 2022A&A...662A..56K}. For example, in our 16 $M_{\rm \odot}$ model with solar metallicity, the mass of the H envelope left after the mass transfer is $\sim$ 1.3 $M_{\rm \odot}$, whereas in the case of LMC it is $\sim$ 1.6 \msun. This preservation of a larger envelope in lower metallicity models  arises from the diminishing efficiency of the stripping procedure under conditions of lower metallicty, as a consequence of the decreased opacity observed in the outer layers. When opacity is decreased, a larger portion of the hydrogen-rich envelope stays within the Roche lobe subsequent to the detachment of the stars. When the rates of wind mass loss are lower, either due to intrinsic factors or due to lower metallicity \citep[eg.][]{2020MNRAS.499..873S}, stars lose less mass in the subsequent post-mass transfer evolution, such that there are more likely to retain H even until the end of their evolution and a supernova.

Eventually, after a phase of post-RLOF contraction, the star evolves into a hot, compact helium star, deriving its primary energy from helium fusion through the triple alpha reaction at its core. This stage is the longest lived phase in the lifecycle of a stripped star, spanning almost 10$\%$ of its total life time. It is marked by point G in Fig. \ref{fig:represent}. However, before the star settles down as a helium star, it undergoes a transition phase (E-F). The stars in this phase are somewhat cooler and more puffed-up than the hot and compact He stars. Due to this, they are visible in the optical spectrum. We denote this stage as puffed-up stripped star (PSS) phase throughout this paper and discuss it in more detail in the following section.

\subsection{A closer look at the contraction phase}  \label{sec3.2}
The dynamics of contraction or expansion of a star can be explained through the examination of its gravitational luminosity ($L_{\text{grav}}$) (as mentioned in Sec \ref{sec3.1}). The contribution of $L_{\text{grav}}$ is generally insignificant during phases of hydrostatic and thermal equilibrium (|$L_{\text{grav}}$| / $L_{\text{surf}}$ $\sim$ 0.01 during MS). However, there is a distinct surge in its magnitude towards the end of the main sequence (MS) phase (the point marked B in Fig. \ref{fig:represent} and see Fig. \ref{fig:panel}), where |$L_{\text{grav}}$| / $L_{\text{surf}}$ $\sim$ 0.9 and |$L_{\text{grav}}$| / $L_{\text{H}}$ $\sim$ 8. This phenomenon arises from the depletion of hydrogen in the core (indicated by a dip in the hydrogen burning luminosity), prompting core contraction to restore equilibrium. As the core contracts, the envelope expands, causing a subsequent reduction and eventual inversion of $L_{\text{grav}}$—its value turns negative. The point of highest magnitude in $L_{\text{grav}}$ is achieved as the star enters the phase of mass transfer, characterized by a substantial negative peak (|$L_{\text{grav}}$| / $L_{\text{surf}}$ $\sim$ 3.24 and |$L_{\text{grav}}$| / $L_{\text{H}}$ $\sim$ 0.77). This phase is very brief and aligns with the period of maximum mass transfer rate (top panel (blue) of Fig. \ref{fig:panel}). Following this peak, the rate of mass transfer experiences a gradual decline. By this time, the star is left with a very little hydrogen envelope ($\sim$ 1.6 \msun) as shown in top panel (red) of Fig. \ref{fig:panel}. Already during the mass transfer, the entire star embarks on a contraction phase. This transition is illustrated in the final panel of Fig. \ref{fig:panel}, where we can see that the $L_{\text{grav}}$ becomes positive shortly after point D. Notably, by the time the mass transfer ends, the star is pretty much in thermal equilibrium, evidenced by the drop in |$L_{\text{grav}}$|. During the PSS phase (shown by thick black lines in Fig. \ref{fig:panel}), the quantity  |$L_{\text{grav}}$| / $L_{\text{surf}}$  ranges between $\sim$0.02 and $\sim$0.005, which is small and typical for nuclear-timescale evolutionary phases.

\begin{figure*}
  \begin{minipage}[c]{0.67\textwidth}
    \includegraphics[width=\textwidth]{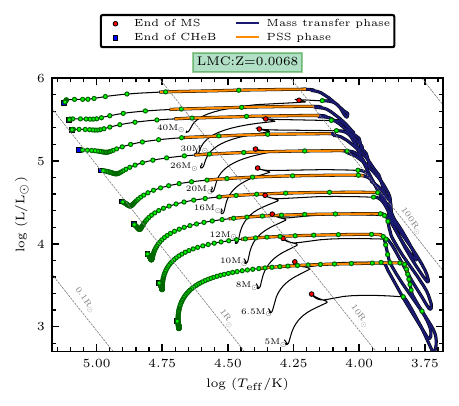}
  \end{minipage}\hfill
  \begin{minipage}[c]{0.3\textwidth}
    \caption{Hertzsprung-Russel Diagram showing evolutionary tracks of the primary star in a binary system with initial masses 5, 6.5, 8, 10, 12, 16. 16.9, 20, 22, 26, 30 and 40 $M_\odot$ at LMC metallicity. The blue lines represent the mass transfer phase and orange lines represnt puffed-up stripped star (PSS) phase. Red dots represents the terminal age main sequence (TAMS) and the blue square represents the end of core Helium burning. The green dots are placed at intervals of 20,000 years strating from RLOF. The clustering of these dots towards the left shows that the star spends a long time at this phase.} \label{fig:tracks}
  \end{minipage}
\end{figure*}

The finding that stripped stars are close to thermal equilibrium right after the end of stable mass transfer has consequences for the timescale of their subsequent contraction. In the existing literature, it has been implied that after detachment stars regain thermal equilibrium only after becoming hot stripped stars, indicating contraction timescales (and the PSS lifetimes) of the order of the thermal timescale of several thousand years \citep[e.g.][]{2017A&A...608A..11G,2020MNRAS.495.2786E,2020A&A...638A..55K}. We find that this is not true. For the example of a 16 $M_{\rm \odot}$ model at LMC metallicity shown in Fig. \ref{fig:represent} and \ref{fig:panel}, the contraction from point E to F lasts for about $\sim6\times10^4$ yr, i.e. at least an order of magnitude longer. This is not isolated example for this particular mass. In Fig. \ref{fig:timescales}, we compare the PSS lifetimes across a wider range of initial primary masses (LMC metallicity) to the thermal timescale $\tau_{th}$ estimated following the expression from \citet{1997PASP..109.1394K}: 
\begin{equation}
    \tau_{th} \approx \frac{GM_d^2/R_d}{L}\simeq 3 \times 10^7 \biggl(\frac{M_d}{M_\odot}\biggl)^{-1/2}\biggl(\frac{R_d}{R_\odot}\biggl)^{3/2} \text{yr}
\end{equation}
This figure illustrates that the PSS phase is 10-100 times longer than the thermal timescale and about 10\% of the He burning lifetime. 
This is in agreement with \cite{2021MNRAS.502.3436E}, who noted a similar ratio of timescales in the low-mass regime of models.

The reason why the star contracts following detachment from point E to G is due to continuous loss of mass from the remaining hydrogen envelope that was left unstripped during the mass transfer. The timescale of contraction and the PSS phase is thus dictated by the rate at which the envelope mass $M_{\rm env}$ decreases. There are two processes that contribute. One is stellar winds that directly strip mass from the envelope. The other is H-shell burning, which gradually grows the mass of the core, reducing $M_{\rm env}$. Perhaps somewhat surprisingly, we find it is the H-shell burning that is the dominant process reducing $M_{\rm env}$ for most initial primary masses (Fig. \ref{fig:percent}). At LMC (SMC) metallicity, the winds begin to dominate only above about 54 M$_\odot$ (63 M$_\odot$) for our particular set of assumptions.

\begin{figure*}
    \centering
`    \includegraphics[width=9cm,height=7cm]{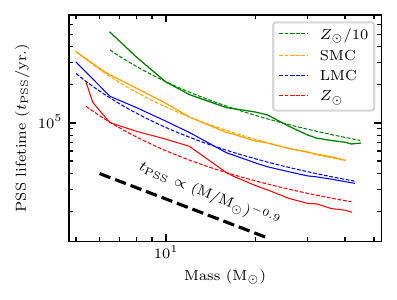}
    \includegraphics[width=9cm,height=7cm]{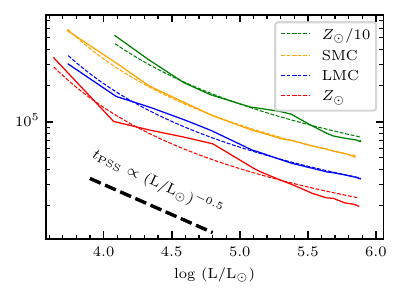}
    \caption{Variation of puffed-up stripped star (PSS) lifetime with initial donor masses (left) and corresponding luminosity (right) at various metallicity (represented by various colours). Dashed lines represent fits for the corresponding data. The bold dashed line guides the eye to illustrate the steepness of the slopes.
}
    \label{fig:fit}
\end{figure*}

The reason why the PSS phase is a transitional stage that is short compared to the He burning lifetime is because the envelope layer left after the mass transfer is thin and gets depleted. Had there been more envelope left after the mass transfer, the contraction PSS phase would last longer. This is the case for instance in the high-semiconvection low-metallicity models of \citet{2022A&A...662A..56K}, where often $\sim25-40$ \% of the initial envelope remains unstripped and the star never contracts to the hot stripped-star stage, with the PSS phase lasting for the entire He burning lifetime. More massive residual envelopes are usually also more H-rich as a result of H/He gradients \citep{2019A&A...625A.132S,2020A&A...638A..55K,2020MNRAS.496.1967K,2020A&A...635A.175H}, depleting the rate at which H-shell burning grows the He core mass and reduced $M_{\rm env}$. To test this and further illustrate this point, we undertake an experiment. We initiate by evolving a binary system until the end of mass transfer. We then stop the evolution of the binary model, extract the donor star, switch off stellar winds, and let the stripped donor evolve further without any wind mass loss. In another experiment, instead of switching off the winds we disable any composition changes due to nuclear burning. The results are shown in Fig.\ref{fig:nowinds_rxn} for the same 16 $M_\odot$ example with LMC composition. In the illustrative model presented in Fig. \ref{fig:represent}, the duration of the phase characterized by a puffed-up stripped star is approximately 0.058 Myr. In contrast, in the scenario where the winds were deactivated, this duration extends to about 0.078 Myr.  On the other hand, disabling compositional changes results in an extended period of approximately 0.27 Myr for the PSS phase. This example illustrates that by lowering the envelope depletion rate we prolong the PSS lifetime and slow down the contraction. In the same time, wind mass loss rate (which is the uncertain factor in stellar models) is subdominant compared to H-shell burning. This suggests that the PSS lifetimes inferred in this work are not significantly affected by uncertainties in the wind mass loss rate of stripped stars (with the exception of very high masses).

\subsection{Stellar tracks}
 In Fig. \ref{fig:tracks}, we present HR diagrams from stellar masses ranging from 5 $M_\odot$ to 40 $M_\odot$ LMC (Z = 0.4$Z_\odot$ =  0.0068) metallicity. The tracks show the evolution of the primary star from ZAMS to end to He burning . We mark the important phases in the evolutionary history as described in the legend.

We find that the duration of the PSS phase depends on the initial mass and the metallicity. In Fig. \ref{fig:fit} (left panel), we plot the lifetime of the puffed-up stripped star as a function of the initial mass (before mass transfer) and for various metallicities. Dashed lines represent numerical fits described by the relation:
\begin{equation}
    T_{\rm PSS}=\biggl(\frac{a_{11}M+a_{21}}{M^2+a_{31}M+a_{41}}\biggl) \biggl(\frac{a_{12}Z+a_{22}}{a_{32}Z+a_{42}}\biggl)
    \label{eqn:mfit}
\end{equation}
where $T_{\rm PSS}$ = $\log (t_{\rm PSS}/\rm yr)$, $M$ = $\log (m/M_\odot)$,  and $t_{\rm PSS}$ is the lifetime of puffed-up stripped star, m is the initial mass and Z is the metallicity. The values of co-eficients $a_{ij}$ are given in Table \ref{tab:mass}. Notably, we find that the duration of the PSS phase $t_{\rm PSS}$ can be well described by two independent factors, on related to mass, the other to metallicity. It can we written as:
\begin{equation}
     \log (t_{\rm PSS} / {\rm yr}) = T_{\rm PSS} = g_m(m)f_m(Z)
\end{equation}

\begin{figure}
    \centering
    \includegraphics[width=9cm,height=6cm]{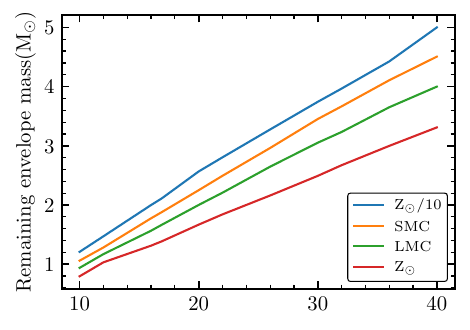}
    \includegraphics[width=9cm,height=7cm]{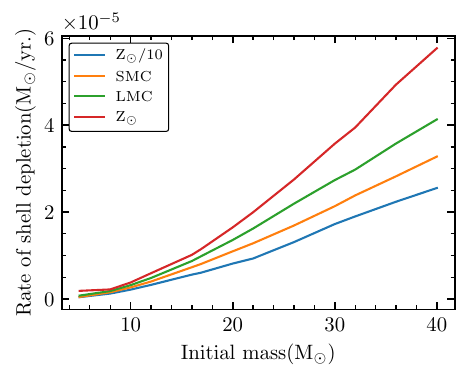}
    \caption{Variation of the remianing envelope mass $M_{\rm env}$ (top) and average rate of $M_{\rm env}$ depletion (bottom) in stripped stars, plotted as a function of the initial mass. The combination of those two factors affects the duration of the PSS phase.}
    \label{fig:rem_env}
\end{figure}

The relationship between the lifetime of a puffed-up stripped star, its mass, and its metallicity can be observed in Fig. \ref{fig:fit}. The decreasing slopes can be explained by considering two key mass- and metallicity-dependent factors: the amount of remaining (hydrogen) envelope mass after mass transfer and the rate of $M_{\rm env}$ depletion due to H shell burning and winds. Fig. \ref{fig:rem_env} illustrates these factors. When we compare models at a constant initial mass, we find that models with lower metallicity tend to retain a larger envelope mass after mass transfer, while models with higher metallicity lose more of their envelopes. Additionally, the rate of shell depletion decreases as we move to lower metallicities. Consequently, models with lower metallicity take a longer time to contract, and they naturally spend more time as PSSs. Conversely, models with higher initial mass, while retaining a larger envelope mass after mass transfer, experience a high rate of shell depletion. This leads to shorter contraction times and, subsequently, shorter PSS lifetimes when compared to models with lower initial mass. 

 \begin{table}[ht!]
     \centering
     \begin{tabular}{c|c c c c}
     \hline \hline
     $a_{ij}$ & i=1 & i=2 &i=3 &i=4 \\
     \hline
     j=1 & 4.57 $\times$ $10^6$ & 2.31  $\times$ $10^6$ & 1.17  $\times$ $10^6$ & 1.69  $\times$ $10^5$\\
     j=2 & 348.54 & 1.27 & 386.68 & 1.14\\
     \hline
     \end{tabular}
     \vspace{3mm}
     \caption{Values of co-efficients $a_{ij}$ in Eqn. \ref{eqn:mfit}}
     \label{tab:mass}
 \end{table}

A closer look at Fig. \ref{fig:tracks} suggests that the luminosity during the PSS phase remains almost constant. Therefore, we could use this luminosity as a proxy for the initial mass to rewrite the fits to $t_{\rm PSS}$ as a function of the luminosity of the PSS phase. To this end we also present corresponding fits following equation similar to Eqn. \ref{eqn:mfit}. Fig\ref{fig:fit} (right panel), shows the lifetime of PSSs as a function of the PSS luminosity at different metallicities. The luminosity is represented by the average the luminosity over the cool stripped star phase (time-weighted). The solid lines represent the lifetime of PSS star from the binary MESA models. Dashed lines represent numerical fits described by the relation:

\begin{equation}\label{eqn:lfit}
     T_{\rm PSS}=\biggl(\frac{b_{11}l+b_{21}}{l^2+ b_{31}l+b_{41}}\biggl)\biggl(\frac{b_{12}Z+b_{22}}{b_{32}Z+b_{42}}\biggl)
\end{equation}
where $T_{\rm PSS}$ = $\log (t_{\rm PSS}/\rm yr)$, $l$ = $\log (L/L_\odot)$,  and $t_{\rm PSS}$,  L is the corresponding luminosity and Z is the metallicity. The values of co-efficient $b_{ij}$ are given in Table \ref{tab:lumin}. 

 \begin{table}[ht!]
     \centering
     \begin{tabular}{c|c c c c}
     \hline \hline
     $b_{ij}$ & i=1 & i=2 &i=3 &i=4 \\
     \hline
     j=1 & 344.76& $-$483.16 & 78.33 & $-$167.01\\
     j=2 & 386.92 & 1.64 & 434.17 & 1.47\\
     \hline
     \end{tabular}
     \vspace{3mm}
     \caption{Values of co-efficients $b_{ij}$ in Eqn. \ref{eqn:lfit}}
     \label{tab:lumin}
 \end{table}

Similar to Eqn. \ref{eqn:mfit}, this expression of time has two dependences: luminosity and metallicity. It can we written as:
\begin{equation}
      \log (t_{\rm PSS} / {\rm yr}) = T_{\rm PSS} = g_L(L)f_L(Z)
\end{equation}

\section{Discussion} \label{sec4}
\subsection{How many puffed-up stripped stars can we expect in stellar populations?}\label{sec4.1}

The tracks computed in our work show that the lifetimes of puffed-up stripped stars are 10-100 times longer than just short thermal timescales that were sometimes assumed before, averaging 10\% of the H burning lifetime and 1\% of the total lifetime of a star.This is in line with the recent surge of detections of PSS binaries in optical spectroscopic surveys.
In this section, we use our results to make an estimation of how many such binaries one can expect to find in low metallicity environments like the LMC and SMC. The detailed procedure for our estimation is explained below.

\begin{figure*}
    \centering
    \includegraphics[width=9cm]{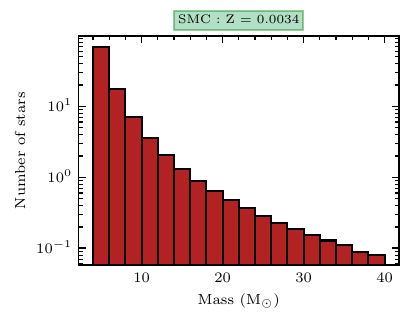}
    \includegraphics[width=9cm]{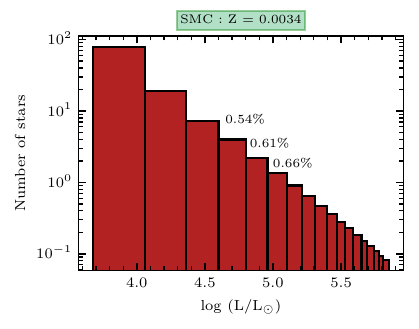}
    \includegraphics[width=9cm]{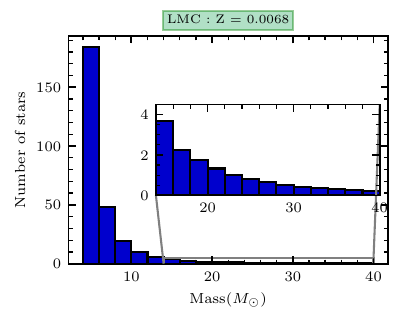}
    \includegraphics[width=9cm]{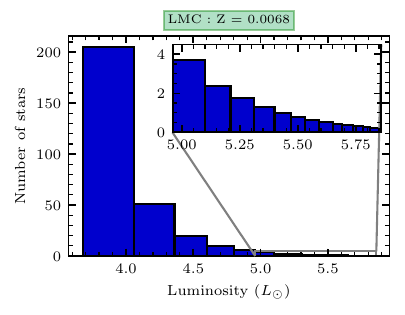}
    \caption{Estimates of the number of stars that can be found at different mass ranges (left) and different luminosity ranges(right) at SMC (upper) and LMC (lower) metallicities. Inset shows a zoomed in version of the ranges which are not clearly visible. Values above the bars in the upper right panel shows the percentage of puffed-up stripped stars relative to the number of all stars in that luminosity bin, most of which are MS, based on \cite{2021A&A...646A.106S} }
    \label{fig:popukation}
\end{figure*}

We assume galaxies with a constant star formation rate (SFR). The SFR of SMC is taken as 0.05 \msun /year \citep{2011ApJ...741...12B,2017MNRAS.466.4540H,2015MNRAS.449..639R, 2018MNRAS.478.5017R} and 0.2 \msun/yr for LMC \citep{2009AJ....138.1243H}. We are interested in finding stars in a specific phase in their evolution, one that lasts $\Delta$t\textsubscript{evol} and begins when the star is t\textsubscript{age} old. In out case $\Delta$t\textsubscript{evol} corresponds the puffed-up stripped star phase. The total mass of stars (M\textsubscript{stars}) formed in that episode of star formation is then SFR(t\textsubscript{age}) $\times$ $\Delta$t\textsubscript{evol}, where SFR(t\textsubscript{age}) is the star formation rate in that galaxy t\textsubscript{age} years ago. If we assume a constant star formation rate, then that means that SFR is independent of t\textsubscript{age}. The M\textsubscript{stars} can be calculated for any of the PSS duration values ($\Delta$t\textsubscript{evol}), assuming a star formation value and history of a particular galaxy. This would correspond to N\textsubscript{stars} = M\textsubscript{stars} / $\langle$ M\textsubscript{star} $\rangle$ stars, where N\textsubscript{stars} is the number of stars and $\langle$ M\textsubscript{star} $\rangle$ is the average mass of stars, given the Initial Mass Function (IMF). We consider the IMF from \cite{2002Sci...295...82K}:
\begin{equation}
    \frac{dN}{dM} = \xi(M) \propto M^{-\alpha_i}
\end{equation}
where $\alpha_1$ = 1.3 for M/\msun $\in$ [0.08, 0.5] and $\alpha_2$ = $\alpha_3$ = 2.3 for M/\msun $\in$ [0.5, 1.0] and [1.0, 150.0] respectively. We normalise $\xi(M)$ to find the constants and from the final value we calculate $\langle$ M\textsubscript{star} $\rangle$:
\begin{equation*}
    \langle M_{\text{star}} \rangle= \int_{0.08}^{150.0} M \, \xi(M) \, dM
\end{equation*}This gives us $\langle M_{\text{star}} \rangle$ as 0.635.

To account for the lifetime of the puffed-up stripped stars derived in Eqn. \ref{eqn:mfit}, we use mass bins at an interval of 2 solar masses ranging from 5 $M_{\rm \odot}$ to 40 $M_{\rm \odot}$. $\Delta$t\textsubscript{evol} corresponds to a particular mass bin. This means that we are only interested in what fraction of all stars ($\text{frac}_M$) formed during a star formation period that produced $N_{\text{stars}}$ in a certain mass bin, for instance, which goes from $M_1$ to $M_2$. We integrate a normalized IMF from $M_1$ to $M_2$ to get $\text{frac}_M$. $N_{\text{stars\_of\_interest}} = \text{frac}_M \times N_{\text{stars}}$ gives us the number of stars (in singles, binaries, in triples) with the mass from the mass bin and formed at such a time period ago that right now we would see them as puffed-up stripped stars. We also assume that the IMF describes the distribution of primaries in multiple systems. So coupled with an assumption of an initial binary fraction of 100\%, $N_{\text{stars\_of\_interest}}$ roughly gives us the number of binary systems that formed with primaries in our mass bin, $t_{\text{age}}$ years ago, during a period of $\Delta t_{\text{evol}}$ years.

In this process, we have considered all binaries without any constraints on mass ratios or periods. However, throughout this work, we specifically consider case B binaries and therefore need to put a constraint on the mass ratio and orbital periods. To this end, we consider distribution functions for orbital periods (P) and mass ratios (q) following \cite{2015ApJ...814...58D}. These distributions were motivated from the spectroscopic observations of massive stars by \cite{2012Sci...337..444S}. The distributions are:
\begin{equation}
    f_P(\log P) \propto (\log P)^\pi, \hspace{6mm} \text{for} \log P \in \text{[0.15, 5.5]}
\end{equation}
where $\pi$ = 0.5
\begin{equation}
    f_q(q) \propto q^\kappa, \hspace{6mm} \text{for q}  \in \text{[0.1, 1.0]}
\end{equation}
where $\kappa$ = $-$ 0.1 $\pm$ 0.6
The range for $\log P$ is chosen from Fig. C.2 of \cite{2020A&A...638A..55K} such that it corresponds to case B mass transfer evolution.
Here, we focus on stable mass transfer evolution, meaning we want to exclude cases that are likely to experience a common-envelope phase. For simplicity, we assume that mass transfer remains stale for $q>0.33$ (radiative-envelope donors) and $q>0.66$ (convective-envelope donors).

Using this method and the underlying assumptions, we were able to estimate the number of puffed-up stripped stars of different masses and luminosities in the environments of the SMC and LMC. Fig. \ref{fig:popukation} shows the results obtained. As can be seen, such stars are more common in the lower (initial) mass and luminosity regime. This would explain the increasing number of low mass stars observed compared to their massive counterparts. Further, as PSS binaries are expected to overlap with the MS, we compare their expected numbers to the number of all MS stars, taking the SMC environment as an example and following \citet[][see their Fig 3]{2021A&A...646A.106S}.
The authors count $\sim$732 stars in the luminosity (log(L/\lsun)) range 4.6 - 4.8, most of which are MS stars. Our estimation of PSS yields $\sim$4 stars in that range, which is about 0.54\% of MS stars. Similarly, in the luminosity (log(L/\lsun)) range 4.8 - 5.1 and 5.1 - 5.3, this number is 0.61\% and 0.66\% respectively. We conclude that PSS stars may constitute about 0.5-0.7\% of the MS population. This may be expected given that the PSS lifetime is about 1-2\% of the entire stellar lifetime and that not all the binaries are going to evolve through stable mass transfer and experience a PSS phase. Although estimated for the SMC, we expect this fraction to be similar in the LMC and the Milky Way environments. This implies that there are still tens of PSS stars disguised as MS stars in the current and upcoming optical surveys.

\subsection{When does the PSS phase end?}
\label{sec4.2}
The transition between  \lq puffed-up' and \lq hot and compact' stripped stars is not easily defined. A useful distinction between different types of stripped-star binaries can be made based on their spectral features, varying as a function of the stripped-star surface temperature and its relative contribution to the total flux of the binary (\citealp{2023arXiv230700061D}, \citealp{2023arXiv230700074G}). In this work, our goal was to adopt an approximate criteria, sufficiently robust to provide at least an order-of-magnitude estimate for the number of PSS in a stellar population. To this end, we assumed that the PSS begins at the end of mass transfer and continues until the effective temperature of the stripped star becomes hotter than its ZAMS temperature plus 0.1 dex. This was motivated by focusing on those stripped stars that in the HR diagram overlap with the MS and could therefore be hiding among the \lq normal' OB stars. Finding candidates for post-interaction systems containing such PS stars could be done by selecting OB stars that stand out in terms of their low surface gravity (or high luminosity to mass ratio, if the luminosity is known) and high nitrogen enrichment, while also usually being slow rotators \citep{2022A&A...662A..56K}.

\begin{figure}[ht]
    \centering
    \includegraphics[width=9cm]{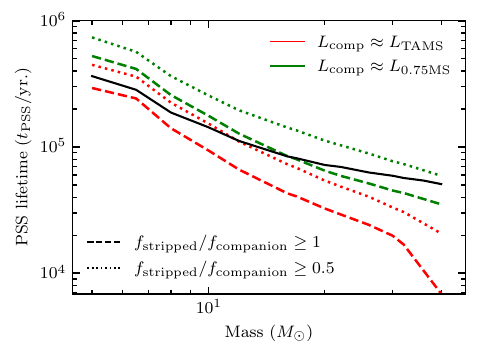}
    \caption{Duration of the puffed-up stripped star (PSS) stage when the flux of the PS star in the optical F450W filter is at least equal (dashed) or half (dotted) the flux of its companion, for two assumed companion lumonosities (red and green, see text), compared to our default PSS stage criteria based on the HR diagram position (black, solid). These are calculated at SMC metallicity.}
    \label{fig:comp_rat}
\end{figure}

In practice, the position of the binary in the HR diagram and the prominence of those spectral stripped-star features will depend on the relative contribution of both binary components to the combined spectra. An extensive discussion of observability of PS star binaries would require exploring a suite of composite combine spectra for a range of different companions, which is beyond the scope of this paper. Instead, we check how the duration of the observable PSS phase would vary assuming that it lasts for as long as the relative flux of the PS star to its companion in the F450W optical filter of HST WFPC2 camera is above a certain threshold. For this experiment, we assume the HR diagram position of the companion is either that of the primary at TAMS or that of the primary at 75\% of the MS lifetime (red and green respectively in Fig. \ref{fig:comp_rat}). The figure illustrates that our default criteria (solid black line) yields PSS lifetimes are comparable to those assuming $f_{\rm stripped}/f_{\rm companion} > 1$ for companions at 0.75 of the MS or to $f_{\rm stripped}/f_{\rm companion} > 0.5$ for companions at the TAMS luminosity of the primary. More luminous companions are also possible: assuming a companion brighter by another factor of 0.2 dex with respect to primary TAMS (such that its bolometric luminosity is similar to that of a stripped star) would mean that the dotted lines in Fig. \ref{fig:comp_rat} correspond to $f_{\rm stripped}/f_{\rm companion} > 0.3$. However, even with a companion $\sim3$ times brighter than the PS star it may still be possible to detect the stripped star characteristics in the combined spectra if they companion is fast rotating and its lines broadened, as is usually the case in post-interaction binaries detected to date. 
Thus, we expect our default criteria for locating the PS phase to be accurate enough for the purpose of order-of-magnitude estimation of the number of PS star binaries in stellar populations. Notably, such systems may also ultimately be identified and spectrally disentangled in cases when it is the companion that strongly outshines the stripped star, provided that the stripped star contributes at least $\sim10-20\%$ of the combined flux \citep{2023MNRAS.521.3162S}.

\subsection{Puffed-up stripped stars after core He depletion}
After the PSS phase, the star enters the phase of core He burning. Due to the energy produced by He burning, the contraction is halted. At the end of core-He burning, upon central He depletion, the carbon-oxygen core begins to contract again and the remaining envelope expands. In some cases this leads to a substantial expansion of the entire star again (even up to hundreds of $R_{\odot}$), such that the star begins to appear as a PS star again \citep{2020A&A...637A...6L}. We follow the behavior of models until the end of central C burning, after which point there is generally not enough time for any significant change in the radius of the star before the core collapse.

The post-core He burning expansion takes place in all our models below $\sim$ 10 \msun.
Going to higher masses, the degree of late expansion becomes metallicity dependent: high metallicity massive models lose too much H/He envelope during core-He burning, such that they remain compact until the core collapse. For example, in our representative models of 16  $M_{\rm \odot}$ (Fig. \ref{fig:represent}), the model with LMC metallicity expands upto $\sim$ 1 $R_{\rm \odot}$; whereas the model with lower metallicity ($Z_{\rm \odot}/10$) expands upto $\sim$ 100 $R_{\rm \odot}$. 
The metallicity effects on post-core He burning expansion are discussed in detail by \cite{2020A&A...637A...6L}.

As a result of this expansion, stars exhibiting similar characteristics to puffed-up stripped stars observable in the HR diagram.  Notably, PS stars at this expanding stage are more luminous compared to PS stars contracting after the end of mass transfer, having thus even higher L/M ratio (which might be used to distinguish expanding PS stars from the contracting ones), given sufficiently accurate determination of atmospheric parameters. These particular stars were omitted from our star count estimation and our lifetimes are for the contracting ones only. This is justified because the duration of the expansion phase (which coincides with the temperature range of the PSS phase) is notably shorter (by a factor of $\sim 10$), making it considerably challenging to detect. For example in our 16 $M_{\rm \odot}$ model at $Z_{\rm \odot}/10$ metallicty (right panel of Fig. \ref{fig:represent}), the time spent in that phase (considering the phase lasts from about H to I) is 12.46 kyr compared to our PSS lifetime of 126.55 kyr.

\subsection{Factors affecting the contraction time}
The computed PSS lifetimes and the estimated numbers of PSS binaries expected in a stellar population are affected by several uncertainties. Here we summarize the expected impact of the main uncertainties (winds, internal mixing, and the single choice of orbital period and mass ratio), with the more detailed investigation given in the appendix.

To estimate the number of PSS binaries in the SMC and LMC environments (Sec.~\ref{sec4.1}), we relied on PSS lifetimes derived from our grids of binary-evolution models. Those models were computed for a single choice of orbital period ($P_{\rm ini} = 70$ days) and mass ratio ($q_{\rm ini} = 0.6$), whereas in a real population binaries with other periods and mass ratios clearly exist. In App.~\ref{A1} we explore how the PSS lifetime changes for varying $P_{\rm ini}$ and $q_{\rm ini}$ (\ref{fig:orbitavar} and \ref{fig:ratio}). In general, longer periods and more equal mass ratios lead to more of the envelope being unstripped during the mass transfer, and consequently to longer PSS lifetimes. In the case of mass ratio the effect is quite small, with the PSS lifetime changing by less than 15\% for $q_{\rm ini}$ ranging from 0.4 to 1.0. In the case of the varying orbital period, the effect is more significant. We find that in the range of periods from $\sim10$ to$\sim1000$ days (i.e. approximately the entire case B mass transfer range), the PSS lifetime vary by up to $\sim2.5$ with respect to our reference model at $70$ days, increasing with $P_{\rm ini}$. As most of the parameter space for case B mass transfer are periods $P_{\rm ini} > 70$d,  the estimated number of PSS stars in Sec.~\ref{sec4.1} (based only case B evolution) is thus likely underestimated by a factor $\sim1-2$ due to our orbital period choice. We also note that the PSS lifetimes originating from case A mass transfer are a factor of a few shorter than those from case B evolution.

In terms of uncertainties due to input physics of stellar models, evolution of stripped stars is particularly sensitive to the assumed wind mass loss rates as well as efficiency of internal chemical mixing. Winds directly affect how quickly the residual unstripped envelope is depleted, which is what determines the PSS lifetime. In App.~\ref{appB} we and Fig.~\ref{fig:winds} we show that enhancing the winds by a factor of 10 leads to PSS lifetimes shorter by a factor of $\sim2.5$ ($\sim 5$) for a stripped star originating from a $16 M_{\odot}$ ($30 M_{\odot}$), respectively. The mass dependence comes from the fact that the lower the mass, the less significant is the contribution of winds relative to H-shell burning in depleting the H-rich envelope. Decreasing the strength of winds with respect to our reference model has minimal effect on the PSS lifetime, unless very high mass stars are considered  (see Sec. \ref{sec3.2} and Fig. \ref{fig:percent}). Internal mixing, on the other hand, plays a role because it affects the chemical gradient of bottom envelope layers of giants, which in turn has an impact on when the binary detaches from mass transfer and the residual amount of unstripped envelope \citep{2019A&A...625A.132S,2020A&A...638A..55K,2020MNRAS.496.1967K,2020A&A...635A.175H}. In addition, the extend of core-overshooting and semiconvective mixing has strong effect on the post-MS expansion of massive low-metallicity stars \citep{1985A&A...145..179L,1991A&A...252..669L,1995A&A...295..685L,2001A&A...373..555M}, which in turn can sometimes lead to the phenomenon of partial-envelope stripping \citep{2022A&A...662A..56K}: case B mass transfer from core-He burning giants leading to residual unstripped envelopes so massive, that the stripped star remains puffed-up and never contracts to become UV-bright and compact. Such cases would lead to PSS lifetimes equaling core-He burning lifetimes, i.e. $\sim10$ times longer than our reference models, see App.~\ref{appC} and Fig. \ref{fig:semiconv}. Such long-lived PSS are restricted to high masses $\gtrsim 15-20 M_{\odot}$ (depending on overshooting) and do not affect the core of our analysis.

\section{Summary and conclusions} \label{sec5}

In this paper, we have studied the post-mass transfer evolution of stars that have undergone stripping in binary systems. We put specific focus on understanding puffed-up stripped stars (PSS), i.e. stars that have lost most of their hydrogen envelope but the remaining envelope is puffed-up, such that the stripped star is as large and cool as normal MS stars or event post-MS giants. To this end, we employed the stellar evolution code MESA code to calculate grids of binary models at four different metallicities (solar, LMC, SMC and 1/10th solar) encompassing a wide range of initial primary masses (from 5M$_\odot$ to 40M$_\odot$). Our principal emphasis revolved around the post-mass transfer evolution of the primary star. For simplicity, the secondary star was treated as point mass. For most of the models, we maintained a consistent set of parameters, initiating the binary systems with an initial orbital period P\textsubscript{ini} = 70 days and an initial mass ratio $q_{\rm ini}$ = 0.6. However, we also conducted exploratory investigations into how variations in P\textsubscript{ini}, $q_{\rm ini}$, wind mass loss, and semiconvective mixing might affect the outcomes in select cases. Our conclusions can be summarized as follows.

\begin{enumerate}
    \item After the end of mass transfer, the primary star (mass loser) initiates a process of contraction to eventually become a hot and compact ($\lesssim 1 \rm R_{\odot}$) stripped star, observable in UV. During contraction it is observable as a much cooler and larger puffed-up stripped (PS) star. The contraction is a transitory phase but is is much longer than the thermal timescale of the primary (by $\sim 2$ orders of magnitude), with PSS lifetimes typically equalling about 1\% of the total stellar lifetime.

    \item Thermal equilibrium is restored very soon after the end of mass transfer and maintained throughout the PSS phase, with negligible contribution from contraction to the total energy budget of the star $\lesssim 1\%$ (Fig. \ref{fig:panel}). The main energy source in PS stars is H shell burning occurring in the residual envelope, unstripped during the mass transfer.

 \item The reason why stripped stars continuously contract after the end of mass transfer is because they continue to lose mass from their residual hydrogen envelope. PS star lifetimes and the pace of contraction are determined by how quickly the residual envelope is depleted. The two agents at play are stellar winds, that directly strip the envelope, and H-shell burning, that grows the He core at an expense of the H envelope. We find that H-shell burning is the dominant effect for initial masses even up to $\lesssim 50 \rm M_{\odot}$, see Fig \ref{fig:percent} (e.g. H-shell burning is responsible for $\approx 90\%$ of the envelope mass loss in PS stars from $\sim10-15 \rm M_{\odot}$ primaries at LMC metallicity).

 \item Uncertainties in wind-mass loss rates of stripped stars have a non-trivial effect on the duration of the PSS phase. Our default assumption of winds follows \cite{2000A&A...360..227N}. Assuming winds stronger by a factor of 10x leads to PSS phase shorter by a factor of $\sim2.5$ ($\sim3.2$) for the initial primary mass $16 M_{\odot}$ ($30 M_{\odot}$), respectively. However, assuming weaker winds has much smaller effect on the PS lifetimes (Fig. \ref{fig:winds}): this is because most of the residual envelope is depleted through H-shell burning anyway.

    \item The duration of the PSS phase depends on the remaining (hydrogen) envelope mass after mass transfer and the rate of envelope mass depletion. The duration decreases with increasing mass (at constant metallicity) due to high rate of envelope mass depletion. With decreasing metallicity (at constant initial mass), the PSS lifetimes increase thanks to higher remaining envelope masses, prolonging the contraction. Similar effect can be observed with orbital periods where the PSS lifetime increases with larger orbital periods and also with initial mass ratio, where greater mass ratio leads to longer PSS duration.

    \item Based on a simple population model, we predicted that about $\sim100$ (in the SMC) and nearly $\sim280$ (in the LMC) PS stars exist in the luminosity range ${\rm log} (L/L_{\odot}) > 3.7$ (corresponding to the initial mass range 5-40 M$_\odot$). We estimate that 0.5-0.7\% of the MS stars in that luminosity range are in fact PS stars in post-interaction binaries, suggesting that many such systems are awaiting detection. We discuss several uncertainties and simplifications of the above estimate, deeming it likely robust within the factor of $\sim 5$.

\end{enumerate}

\begin{acknowledgements}
It is a pleasure to acknowledge valuable inputs and discussions with Julia Bodensteiner. DD would like to thank Projjwal Banerjee for his advice and comments regarding the paper. 
JK acknowledges support from an ESO Fellowship. 
\end{acknowledgements}

\appendix
\section{Effect of orbital period and mass ratio on the lifetime of a puffed-up stripped star} \label{A1}
In this section, we investigate how the orbital period and mass ratio influence the post-interaction contraction thereby the time period of the bloated stripped star phase. We take models with initial stellar mass 16 M$_\odot$ and 30 M$_\odot$ at LMC metallicity (Z = 0.0068) and vary their orbital periods and mass ratios. 

\subsection{Orbital periods}
\begin{figure}[ht!]
    \centering
    \includegraphics[width=9cm]{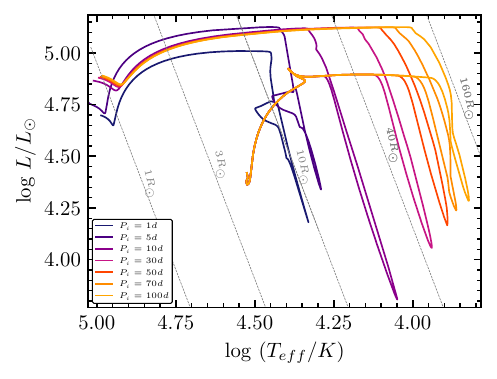}
    \caption{Evolutionary tracks of 16M$_\odot$ models at LMC metallicity (with varied initial orbital periods ($P_{ini}$)) on the HR diagrams. Various colours indicate the various values of $P_{ini}$ describes in the legend.}
    \label{fig:orbitHR}
\end{figure}

In \ref{fig:orbitHR}, we have plotted the full stellar evolution tracks at various orbital period ranging from 1 day to 100 days. The models with periods 1 day and 5 days undergoes mass transfer still when in main sequence (case A). Since our focus is on case B mass tranfer, we do not consider them in our discussion.

\begin{figure}[ht!]
    \centering
    \includegraphics[width=9cm]{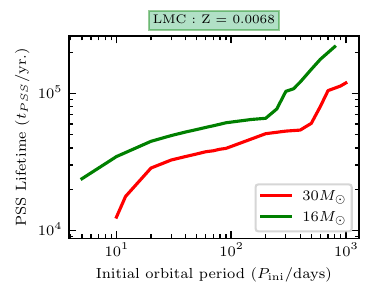}
    \caption{Variation of initial orbital period $P_{ini}$ with puffed-up stripped star (PSS) lifetime of 16 and 30 M$_\odot$ models at LMC metallicity.}
    \label{fig:orbitavar}
\end{figure}

Fig. \ref{fig:orbitavar} shows the variation of the puffed-up stripped star (PSS) lifetime with various orbital period. It can be seen that the PSS lifetime increases with increasing orbital periods. This happens because for longer orbital periods, we have wider separation and thus bigger roche lobes. With a larger Roche lobe, the primary star detaches from its companion sooner compared to a scenario with a smaller Roche lobe. Consequently, stars with longer orbital periods experience less efficient mass stripping, retaining more of their envelope mass. As explained in Section \ref{sec3.2}, the envelope mass of a star is a pivotal factor in determining the duration of its contraction phase. In this context, the increased envelope mass for stars with longer orbital periods leads to a slower contraction process. This phenomenon contributes to the observed extended lifetime of puffed-up stripped stars in such systems.

\subsection{Mass ratio}
\begin{figure}[ht!]
    \centering
    \includegraphics[width=9cm]{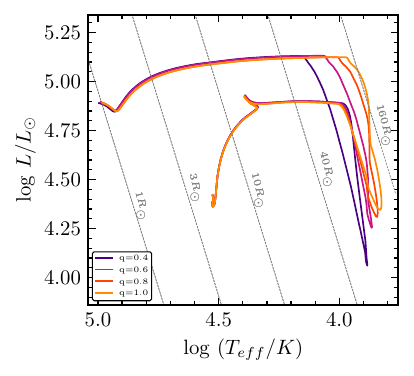}
    \caption{Evolutionary tracks of 16M$_\odot$ models at LMC metallicity (with varied initial mass ratio ($q_{\rm ini}$)) on the HR diagram. Various colours indicate various mass ratios describes in the legend. }
    \label{fig:mratio}
\end{figure}

Fig. \ref{fig:mratio} shows stellar tracks at various mass ratio ($q=M_2/M_1$) ranging from 0.4 to 1.0. Our default mass ratio used is 0.6. We consider all the models here as all of them undergoes case B mass transfer

\begin{figure}[ht!]
    \centering
    \includegraphics[width=9cm]{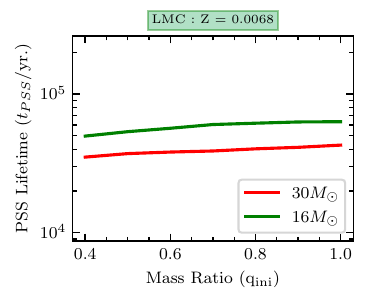}
    \caption{Variation of initial mass ratio ($q_{\rm ini}$) with puffed-up stripped star (PSS) lifetime of 16 and 30 M$_\odot$ models at LMC metallicity.}
    \label{fig:ratio}
\end{figure}

The influence of mass ratio on the orbital period is illustrated in Fig. \ref{fig:ratio}. An increase in mass ratios correlates with an extended lifetime of PSS. This phenomenon arises from the transfer of mass from a more massive star to a less massive companion. Consequently, the orbital separation decreases to conserve angular momentum. The mass of the primary star decreased, reaching a critical point where the mass ratio undergoes reversal, leading to an increase in orbital separation. As the mass ratio approaches unity, the orbit could not shrink at all because due to the rapid reversal of the mass ratio, resulting in a substantial separation between the stars. This expansive separation gives rise to a larger Roche lobe. As explained in the previous section, a larger Roche lobe contributes to inefficient mass stripping, allowing the retention of more envelope mass resulting in longer contraction time

\section{Effect of winds on the lifetime of a puffed-up stripped star} \label{appB}

In the previous section, we discussed how the primary factor influencing the contraction of a puffed-up stripped star is the process of shell hydrogen burning. However, above a certain mass, stellar winds also play a significant role in determining the rate of contraction. This limit is different for different metallicities. For SMC, it is 54\msun and for LMC it is 63\msun. Above this limit, winds can have an impact on the overall lifetime of the PSS.

\begin{figure}[ht!]
    \centering
    \includegraphics[width=9cm]{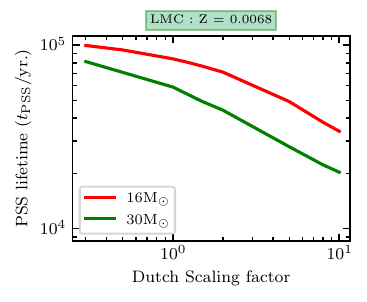}
    \caption{Variation of the puffed-up stripped star (PSS) lifetime with wind strength for 16 and 30 M$_\odot$ at LMC metallicity.  }
    \label{fig:winds}
\end{figure}

There is ongoing debate and uncertainty surrounding the mass-loss rates of stripped stars \citep{2015PASA...32...15Y,2017A&A...608A..11G, 2018A&A...615A..78G}. These rates are generally calculated from the Wolf-Rayet scheme of \cite{2000A&A...360..227N}. The effect of the winds is somewhat less for lower metallictiy \citep{2005A&A...442..587V}. This relation with metallicity exerts an impact on the mass of the residual H envelope. In the context of puffed-up stripped stars, it is important to recognize that the lifetime of a puffed-up stripped star depends on how much envelope mass remains and the nuclear burning rate of the shell. The amount of mass that can be carried away by these winds also plays a role in shaping the evolution and, by extension, its lifetime.

We conducted experiments involving variations in the wind mass loss rates to investigate their impact on the PSS lifetime. Figure \ref{fig:winds} illustrates the relationship between different wind rates and the PSS lifetime. Our findings indicate that as the wind mass loss rates increase, the PSS lifetime decreases. This phenomenon occurs because more vigorous winds result in a less massive envelope surrounding the star, which in turn depletes more rapidly than it would with a larger envelope.

However, it is important to note that the effect of wind mass loss rates on PSS lifetime is not particularly pronounced. The overall difference in PSS lifetime due to these variations falls within a relatively narrow range, typically within half an order of magnitude.

\section{Effect of semiconvection on the lifetime of a puffed-up stripped star}\label{appC}
Semiconvection is an important factor that particularly affects the evolution of post-MS donors. Models with less efficient semiconvection can rapidly expand during the HG phase all the way until the red supergiant stage and be subject to full envelope stripping through thermal-timescale mass transfer. However, models with high semiconvection variant undergo partial envelope stripping and undergo a slow (nuclear-timescale) mass transfer \citep{2020A&A...638A..55K, 2022A&A...662A..56K}. In our models, we have considered less effective semiconvection so as to have a classical case B mass transfer. However, we also explore the case with efficient semiconvection. 

\begin{figure}[ht!]
    \centering
    \includegraphics[width=9cm]{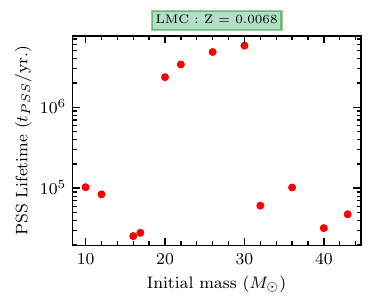}
    \caption{Variation of the puffed-up stripped star (PSS) lifetime with initial mass with a high semiconvection variant ($\alpha_{SC}$=100) at LMC metallicity. }
    \label{fig:semiconv}
\end{figure}


Fig. \ref{fig:semiconv} is a variation of the PSS lifetime with high semiconvection ($\alpha_{SC}$ = 100). This plot shows a strongly different behavior. This is because in this model variation some of the mass transfer sequences are initiated when the donor is already a core-He burning star (the thermal-timescale core contraction has ended before the RLOF). This happens in models in which massive stars transition to the core-He burning while still relatively compact blue supergiants. Mass transfer from such donors was shown to lead to partial envelope stripping, with higher remaining envelope masses, such that post-interaction mass losers may never contract to become hot stripped star and instead live as puffed-up (partially) stripped stars for the entire remaining duration of their core He burning lifetime \citep{2022A&A...662A..56K}. Such phenomena was found in stellar models a long time ago, particularly in the regime of high mass and low metallicity  \citep{1982ApJS...49..447B,1991A&A...245..548B,1991A&A...252..669L,2013A&A...558A.103G,2014MNRAS.445.4287T,2019A&A...627A..24G,2020A&A...638A..55K}. It is highly sensitive to the degree of internal mixing in the envelope layers just on top of the H burning shell, in particular to semiconvective mixing which affects those regions just after the end of the MS. Enhanced mixing above the hydrogen-burning shell significantly influences the H/He composition gradient within the hydrogen-shell burning region and beyond. This, in turn, plays a decisive role in dictating the size of a star as it progresses towards the nuclear-timescale core helium burning phase, as well as the amount of its envelope that is shed during mass transfer. For a detailed discussion, please refer to Sect. 5.1 and Appendix B of \cite{2020A&A...638A..55K}.

We expect this effect to play a role predominantly at low metallicity (SMC and below) and in a certain mass range that is model dependent but generally above 15\msun. This is evident in our results (Fig. \ref{fig:semiconv}). For models above 20\msun, the PSS lifetime increases dramatically. Similar effect could be played by very efficient roationally-induced chemical mixing \citep{2014A&A...566A..21G}

\section{Additional figures}

\begin{figure}[ht!]
    \centering
    \includegraphics[width=10cm]{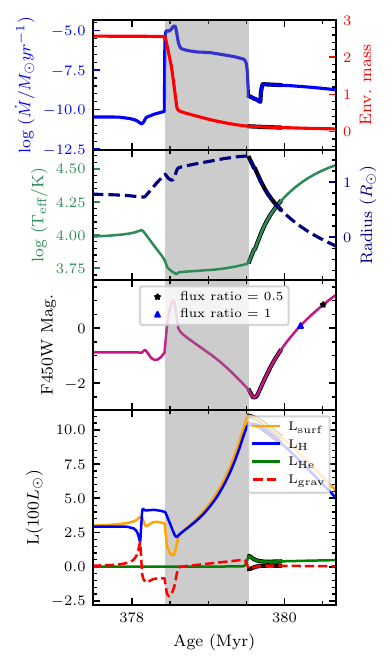}
    \caption{Same as Fig. \ref{fig:panel}, but for 3 $M_\odot$ primary}
    \label{fig:panel6}
\end{figure}

\begin{figure}
    \centering
    \includegraphics[width=10cm]{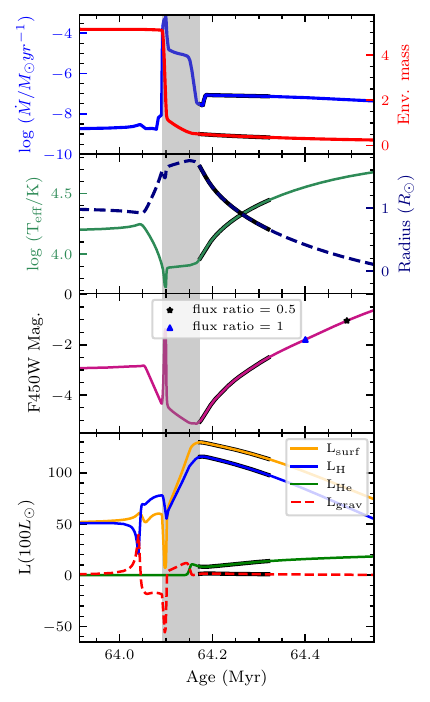}
    \caption{Same as Fig. \ref{fig:panel}, but for a 6.5 M$_\odot$ primary}
    \label{fig:panel3}
\end{figure}

\begin{figure*}
    \centering
    \includegraphics[width=12cm]{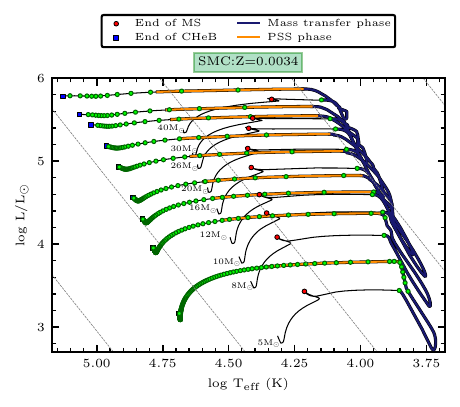}
    \caption{Same as Fig. 6, but for SMC metallicity}
    \label{fig:}
\end{figure*}

\begin{figure*}
    \centering
    \includegraphics[width=12cm]{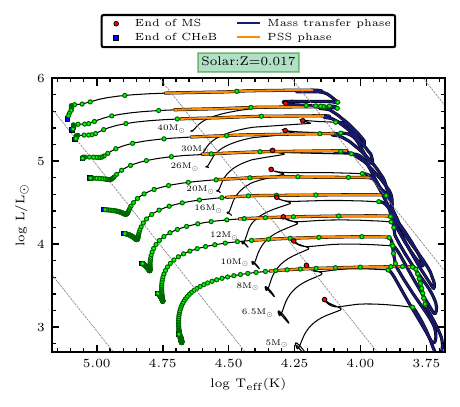}
    \caption{Same as Fig. 6, but for solar metallicity}
    \label{fig:enter-label}
\end{figure*}

\end{document}